\documentclass[preprint,12pt]{aastex}
\usepackage{epsfig}
\usepackage{emulateapj5,apjfonts}
\makeatletter

\newenvironment{inlinefigure}{%
\def\@captype{figure}%
\noindent\begin{minipage}{0.999\linewidth}\begin{center}}
{\end{center}\end{minipage}\smallskip}
\makeatother

\newcommand{\chandra}{{\it Chandra}}
\newcommand{\hst}{{\it HST}}
\newcommand{\lum}{\thinspace\hbox{$\hbox{erg}\thinspace\hbox{s}^{-1}$}}

\def\etal{et al.} 
\def\rd{Di\thinspace Stefano}

\begin{document}

\title{\ \ \ \ \ \ \ \ \ \ \ \ \ \ \ X-ray Point Sources in the Sombrero Galaxy:\hfil\break
Supersoft sources, the globular cluster/LMXB connection, and an overview} 

\author{R. Di\,Stefano\altaffilmark{1,2}, A.K.H. Kong\altaffilmark{1},
M.L. VanDalfsen\altaffilmark{3,4}, 
W.E. Harris\altaffilmark{3}, 
S.S. Murray\altaffilmark{1},
Kisha~M.~Delain\altaffilmark{5}}

\altaffiltext{1}{Harvard-Smithsonian Center for Astrophysics, 60
Garden Street, Cambridge, MA 02138; rd@cfa.harvard.edu}
\altaffiltext{2}{Department of Physics and Astronomy, Tufts
University, Medford, MA 02155}
\altaffiltext{3}{Department of Physics and Astronomy, McMaster
University, 1280 Main Street W, Hamilton, ON, L8S 4M1, Canada}
\altaffiltext{4}{Visiting astronomer, Canada-France-Hawaii Telescope, operated by the National Research Council of Canada, the Centre National de la
Recherche Scientifique de France, and the University of Hawaii}
\altaffiltext{5}{University of Minnesota, Department of Astronomy, 
Minneapolis, MN 55455}

\begin{abstract}
We report on the population of point sources discovered during
an $18.5$ ksec {\it Chandra} ACIS-S observation of the Sombrero
Galaxy. 
We present the luminosity function, the spectra of the 
$6$ brightest sources, consider correlations with
globular clusters (GCs) and with planetary nebulae (PNe),
and study the galaxy's population of SSSs. 
We detected 122 sources, $22$ of them are identified as luminous supersoft X-ray
sources (SSSs). There is an over density of SSSs within $1.5$ kpc
of the 
nucleus, which is itself the brightest X-ray source. 
SSSs are also found in the disk and halo, with one SSS
in a globular cluster (GC).  This source is somewhat harder than most
SSSs; the energy distribution of its photons is consistent
with what is expected from an accreting intermediate mass black
hole. Several sources in Sombrero's halo are good candidates for SSS
models in which the accretor is a nuclear-burning white dwarf.
In total, $32$ X-ray sources are 
associated with GCs. The majority of sources with luminosity 
$> 10^{38}$ erg s$^{-1}$ are in GCs. These results for
M104, an Sa galaxy, are similar to
 what has been found for elliptical galaxies and for the
late-type spiral M31.
We find that those optically bright GCs with X-ray sources house only the brightest
 X-ray sources. We find that, in common with other
galaxies, there appears to be a positive connection between
young (metal-rich) GCs and X-ray sources, but that the brightest X-ray
sources
are equally likely to 
be in metal-poor GCs. 
The luminosity function of X-ray sources in GCs 
has a cut-of near the Eddington luminosity for a $1.4\, M_\odot$
object.
We propose a model which can 
explain the trends seen in the data sets from the Sombrero
and other galaxies. Thermal-time scale mass transfer can occur
in some of the the younger clusters in which the turn-off mass
is slightly greater than $0.8\, M_\odot$;
multiplicity may play a role in some of the most massive clusters; 
accretion from giant stars may be the dominant mechanism
in some older, less massive and less centrally concentrated
clusters. Key elements of the model can be tested.
\end{abstract}

\section{Introduction} 

M104 (NGC 4594), the Sombrero galaxy, is one of the most remarkable
sights in the sky. 
 Beyond its aesthetic appeal, the fact that it
is a bright %(mag) 
Sa galaxy which we view nearly edge on,   % (incl), 
provides insight
into the organization of matter within spiral galaxies,
allowing us to identify objects located
out of the plane of the disk. 
During its second year of operations, {\it Chandra}
conducted an $18.5$ ksec observation
of M104. We report here on the X-ray point sources discovered during
this observation. 

The results are remarkable because, in this 
relatively short observation of a galaxy located
$8.9$ Mpc from us (Ford et al.\ 1996), {\it Chandra} detected 
$122$ point X-ray sources, 
 $22$ of them identified as very soft by  an algorithm 
(\rd\ \& Kong 2003a, b)
developed to search for  
luminous supersoft X-ray sources 
(SSSs). 
These observations provide the first glimpse of SSSs located a
measurable distance away from the plane of the galaxy in which they
reside.

M104 is estimated to contain $\sim 1100$ GCs, and therefore 
provides a good opportunity to study the
relationship between X-ray sources and GCs.
Thirty two of X-ray sources, including $1$ SSS,
are apparently associated with globular
clusters (GCs). In fact, M104 is the first spiral outside the Local Group
 in which 
the connection between GCs and X-ray sources is being studied.
 
M104 is 
a LINER galaxy, and it has been conjectured that it houses an
active galactic nucleus, making the nuclear region a
prime target for X-ray observations.  {\it ROSAT} High-Resolution
Image observations identified $8$ point-like sources,
including one coincident with the nucleus, tentatively
interpreted as a low-luminosity active galactic nucleus (AGN;
Fabbiano \& Juda 1997). {\it Beppo SAX} observations,
coupled with a short ($1.7$ ksec) {\it Chandra} exposure
also support the AGN model, although a superposition 
of sources in the nuclear region could not be ruled out (Pellegrini
et al. 2002).    
The diffuse emission is also being studied (Forman et al. 2003; 
see Delain et al. 2001, which also includes a preliminary analysis
of the point sources).  

The focus of this paper is on the rich point source population
revealed by the unique angular resolution of {\it Chandra}.
In \S 2 we discuss the observations and data reduction,
focusing in \S 3 on an overview of the source properties: 
IDs with GCs and planetary nebulae, 
the
spectra of the nuclear source and of the
$6$ brightest non-nuclear sources, the  
luminosity function, and color-color diagram. 
Our main science results are on SSSs, discussed in \S 4, and 
on the connection between GCs and low-mass X-ray binaries (LMXBs),
discussed in \S 5. \S 6 highlights our conclusions and discusses prospects
for future observations.

\section{Observations and Data Reduction}
M104 was observed by \chandra\ on 31 May 2001 with the Advanced CCD Imaging
Spectrometer (ACIS-S). The effective exposure time was about 18.5
ks. In this paper, we focus only on the data taken with the S3
CCD. The nuclear region of M104 was placed near the aim-point of the
ACIS-S array. All data were telemetered in faint mode and were collected with a frame transfer time of 3.2 s. In
order to reduce the instrumental background, only data with {\it ASCA} grades of 0, 2, 3, 4, and 6 were
included. We ran the detection algorithm in the energy range
$0.1-7$ keV and visually inspected the regions around each source.
 We also inspected the
background count rates from the S1 chip; no significant background
flares were found. The data reduction and analysis was done with CIAO,
Version 2.3 and energy spectra were analyzed with XSPEC, Version 11.2.

Discrete sources in the image were found with WAVDETECT (Freeman et
al. 2002) together with an exposure map. A total of 122 sources were
detected. Source count rates were determined via aperture
photometry. The radius of the aperture was varied 
with average off-axis angle in order to match the
90\% encircled energy function. Background was extracted from an annulus centered on each source. In some cases, for
example, near the nucleus, we modified the extraction region to avoid nearby sources. It was
also necessary to modify the extraction radius for some faint sources that are close to more
luminous sources. Every extraction region was examined carefully in the image. The count rate was
corrected for exposure, background variation, and instrumental PSF.
All sources in the catalog have signal-to-noise ratio (S/N) $> 2.6$
and only 12 sources have S/N $< 3$ with the minimum number of counts 
equal to 7. 

\begin{figure*}
\psfig{file=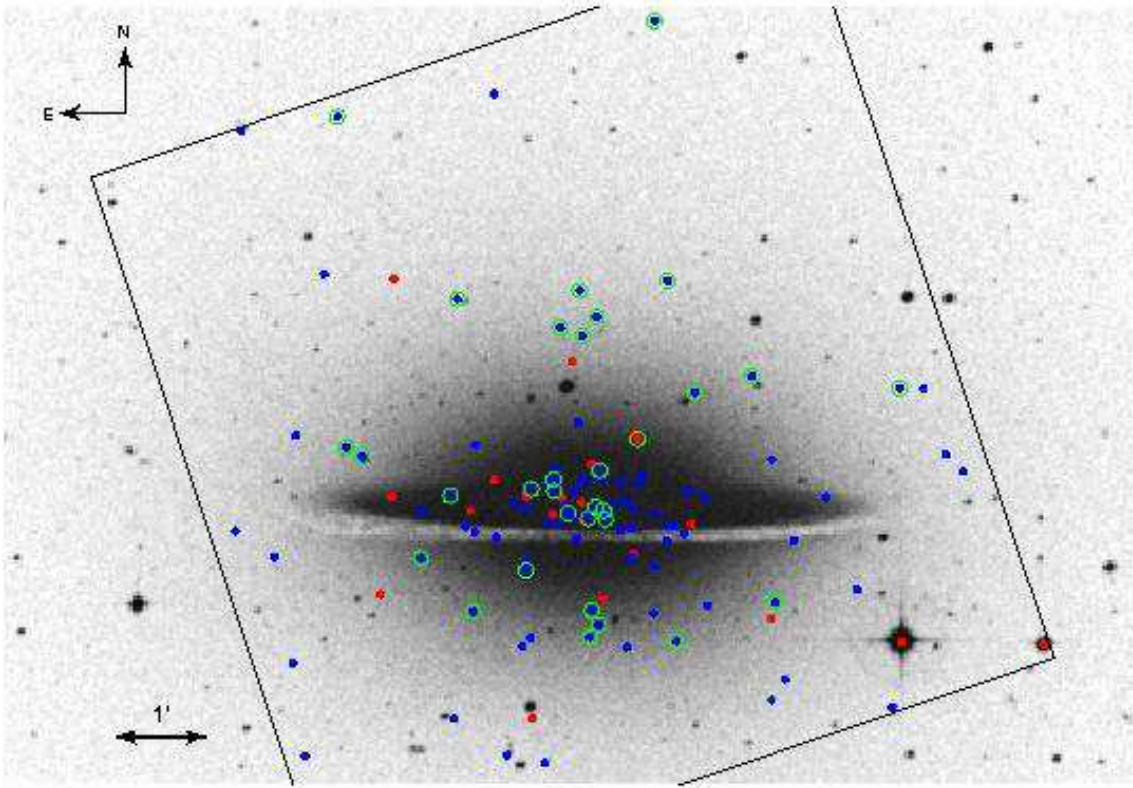,width=6in}
\caption{Detected X-ray sources (blue and red filled circles) overlaid on an optical Digital Sky
Survey image of M104. SSSs are shown as red filled circles while X-ray GCs are
marked with green open circles. To be identified as a match, a GC and an
X-ray source must lie within $1.5''$ of each other; note, however,
that to show the matches clearly, the green circles have a much
larger radius ($5''$).   
The field of view of the S3 CCD is marked as a
black line. North is up, and east is to the left.}
\end{figure*}

\begin{figure*}
\psfig{file=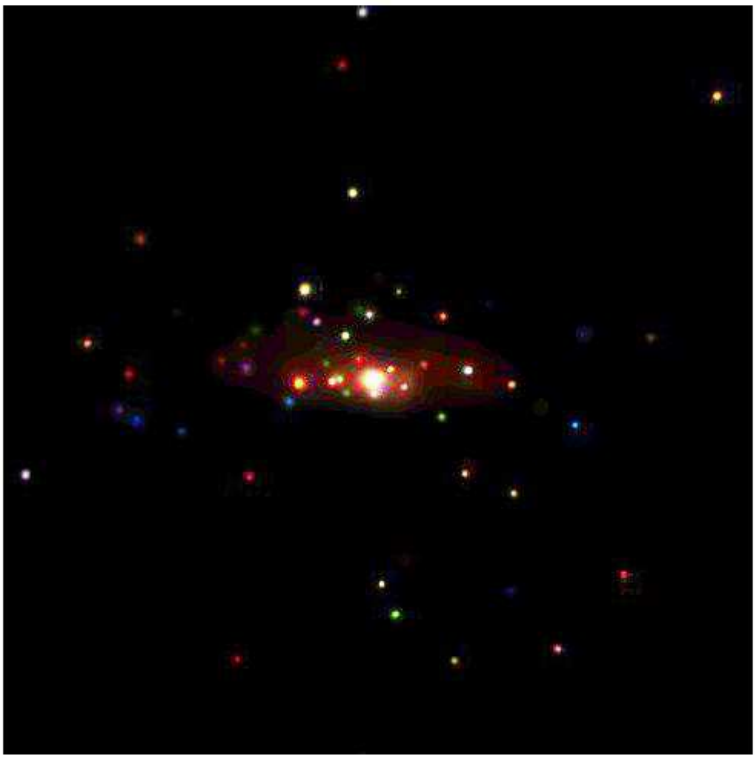,width=4in}
\caption{``True color'' \chandra\ image of the central $4'\times4'$
region of M104. The three energy bands are 0.1--1.1 keV (red), 1.1--2
keV (green), and 2--7 keV (blue). The image has been exposure corrected
and adaptively smoothed. North is up, and east is to the left.
Note the extended diffuse emission near the galaxy's center.
The background subtraction method we applied should minimize the
influence of this diffuse emission on estimates of the 
broadband count rates of point sources; nevertheless, the uncertainty
in the numbers of counts in each band (S, M, and H) is larger than
it would otherwise be for those X-ray sources located in regions with
significant amounts of diffuse emission.  
}
\end{figure*}

Table 1 lists the 122 sources in our catalog, 
sorted in order of increasing right ascension. 
The position of each detected source listed in Table 1 was corrected
for astrometry with the Two Micron All Sky Survey (2MASS) catalog
(Cutri et al. 2000). Within S3, we found six X-ray sources
located within $1''$ of 2MASS stars (see Table 1). We 
calculated the 
average coordinate shift for the six sources and used this to correct the \chandra\
position. The shift in R.A. and Dec. is $0.17''$ and $0.25''$, respectively.
The average displacement between the X-ray and 2MASS
positions after the shift is about $0.15''$.

Some of the sources may be background active galactic nuclei (AGNs). 
We estimated the 
contribution of background unrelated objects by the 
{\it Chandra} Deep Field Surveys (e.g.,
Brandt et al. 2001; Giacconi et al. 2001). 
We found that at $4.5\times10^{37}$\lum\ (the completeness limit), 
fewer than 10 sources are likely to be background objects. 
We expect that only a small fraction of the $\sim 10$ background objects 
are soft enough to be identified as SSSs.  
To quantify this, we have studied data from several
fields observed by {\it Chandra} 
\footnote{http://hea-www.cfa.harvard.edu/CHAMP} 
and find that $1-3$ SSSs unrelated to M104 are likely to be
present in our data; these are most likely to be foreground stars (see
e.g. Di\,Stefano et al. 2003). In fact, 2 foreground stars are 
identified as SSSs. For more details on how the analysis was 
conducted, see \rd\ et al. 2001 and Kong et al. 2002. 

\section{Properties of X-ray Sources}

\subsection{Source Identification}

Observations at optical wavelengths have identified 
globular clusters and planetary nebula in M104 
(VanDalfsen et al. 2001; 
Larsen, Forbes, \& Brodie 2001; Ford et al. 1996). We 
searched for coincidences between these catalogs and our \chandra\
sources.  There are 640 GCs identified from the ground (VanDalfsen et al. 2001)
in the field of view of S3. Moreover, 72 additional GCs were found by
using {\it HST} (Larsen et al. 2001). It is worth noting that the
astrometry of the {\it HST} image (Larsen et al. 2001) was not well
calibrated and we therefore corrected the astrometry by using the
2MASS catalog in our analysis. 
In addition, we also checked the USNO-B1.0 (Monet et
al. 2003) and 2MASS (Cutri et al. 2000) catalogs. We included all the
coincidences in Table 1. Since the astrometry of these catalogs is
better than $1''$, we used a searching radius of $1.5''$ to
search for GCs and planetary nebula. We found 32 X-ray GCs (25 from
the ground and 7 from {\it HST}) out of 742
optically identified GCs within S3, while we did not find X-ray
emitting planetary nebula. The general properties of these X-ray GCs
will be discussed in detail in next subsections. 
There are $3$ ($4$) X-ray GCs that are also
coincident with 2MASS (USNO) sources. 

In order to estimate the accidental correlation rate, we shifted
all the \chandra\ sources by $4''$, $6''$, $8''$ and $10''$ to the north, east, south, west, northeast, northwest, southeast, and southwest, and ran the search for each of the
catalogs. The average accidental matching 
rates for GCs catalog from the ground, GCs catalog from the \hst\, and PN catalog are 2.8, 3.9, and 0.7 respectively. 
We therefore estimate that $22$ of the X-ray sources associated with GCs
identified from ground-based observations and $3$ from {\it HST}
observations represent genuine
matches. 
The relatively higher accidental matching rate for the \hst\ catalog may due to the crowding in the nuclear region of M104. The analyses 
presented in the remainder of the paper apply to only the GCs
observed from the ground.

\subsection{The Nuclear Source}

Pellegrini et al.\, (2002) used a $1.7$ ksec ACIS-S observation to  
study the spectrum of the central source. They found a power-law
fit ($\alpha = 1.5^{+0.4}_{-0.3}$), with 
$N_H= 1.7^{+1.1}_{-0.9} \times 10^{21}$ cm$^{-2}$. The flux was
$1.6 \times 10^{-12}$ erg cm$^{-2}$ s$^{-1}$, which translates to    
$L=1.5 \times 10^{40}$ erg s$^{-1}$ at $8.9$ Mpc. 
During the $18.5$ ksec observation,
photons from the nuclear source were piled up
($\sim 20\%$). By using a pile-up model developed for \chandra\ data (Davis 2001), 
we found a power-law
fit ($\alpha = 2.18^{+0.35}_{-0.35}$), with
$N_H= 2.9^{+0.57}_{-0.57} \times 10^{21}$ cm$^{-2}$; 
assuming the distance of $8.9$ Mpc, $L=2.0 \times 10^{40}$ erg s$^{-1}$.  
The properties of the nuclear source in the second, longer observation
we report on 
are roughly consistent with those described by Pellegrini et al.\, (2002).
They are also consistent with an analysis including
{\it XMM-Newton} data (see Pellegrini et al.\, 2003 for a more 
complete discussion of the nuclear source).

\subsection{Spectral fits of bright sources}

We extracted the energy spectra for the 6 brightest sources 
($> 100$ counts), and
fitted them to simple one-component spectral models including absorbed
power-law, blackbody, disk blackbody, and Raymond-Smith models. 
The background spectrum was extracted from source-free
regions around each source. 
We also corrected for the degradation of soft energy ($< 1$ keV)
sensitivity\footnote{see
http://asc.harvard.edu/ciao/threads/apply\_acisabs/}. Table 2 lists
the spectral fits of these sources. Except for X5 which is a
bright foreground star ($B=10.3$), all sources can be fit with a
single-component model. The spectral shape of X9, X38, X59, and X79 is
consistent with an absorbed power-law model with photon index between
1.1 and 2. These spectra are similar to typical X-ray binaries seen in
our own Galaxy and in other galaxies. Figure 3 shows the energy
spectrum of the brightest point source (excluding the nucleus), X79.
X82 is soft compared with the other sources; the energy spectrum is a
blackbody model with thermal temperature of 0.18 keV (see Table 2 and
Figure 3) and it is also
classified as an SSS (see above). The foreground star, X5, is soft and
requires a Raymond-Smith plus power-law model (see Table 2). 
The values of $N_H$
are generally low and are roughly consistent with the Galactic value
along the direction to M104 ($N_H=3.7\times10^{20}$ cm$^{-2}$).
The 0.3--7 keV
luminosities of these sources (excluding X5) range from $4\times10^{38}$ erg s$^{-1}$
to $1\times10^{39}$ erg s$^{-1}$ at a distance of 8.9 Mpc. 

\vspace{0.6cm}
\begin{inlinefigure}
\rotatebox{-90}{\psfig{file=79spec.ps,height=3in}}
\rotatebox{-90}{\psfig{file=82spec.ps,height=3in}}
\caption{Spectral fit to X79 (power-law model with $N_H=1.1\times10^{21}$
cm$^{-2}$ and $\alpha=2.04$), and X82 (blackbody model with
$N_H=0.85\times10^{-20}$ cm$^{-2}$ and $kT=0.18$ keV).}
\end{inlinefigure}

\subsection{X-ray Luminosity Function}

We constructed the X-ray point source luminosity function (LF). The count rates for all detected
sources were converted into unabsorbed 0.3--7 keV luminosities by
assuming an appropriate spectral model.
For the brightest 6 sources (Table 2), we used the luminosities derived
from the best-fitting spectral model.
In addition, we excluded all the foreground stars (see Table 1). In
Figure 4, we plot the cumulative luminosity functions for all sources
(excluding GCs) and X-ray GCs. 
Because we have eliminated foreground stars and expect that fewer than $10$ 
of the X-ray sources are background AGN (see \S 2), these luminosity functions
should primarily describe the properties of sources associated with M104. 

To estimate the completeness limit, we used similar method by
Kong et al. (2002, 2003a). We computed histograms of
the number of detected sources against the S/N
to examine the completeness limit; the histograms peak at S/N$\sim 3.2$, corresponding
to $4.5\times10^{37}$\lum, and fall off below this. Hence the
luminosity function is complete down to $4.5\times10^{37}$\lum.
It is clear that the two luminosity functions are not well-described by
a simple 
power-law. Instead, they both have a break or cutoff at near
$1-3\times10^{38}$\lum. In addition, power-law models may not be
good descriptions above and/or below the breaks. We nevertheless
use a broken power-law model for each LF, to facilitate comparison
with other treatments (see, e.g., Sarazin et al.\, 2001).  The function is defined
as:

\begin{equation}
\frac{dN}{dL_{38}}=A\left(\frac{L_{38}}{L_{b}}\right)^{-\Gamma}
\end{equation}

where $\Gamma=\Gamma_1$ for $L_{38} \leq L_{b}$, $\Gamma=\Gamma_2$ for
$L_{38} > L_{b}$,
and $L_{38}$ is the 0.3--7 keV X-ray luminosity in units of $10^{38}$\lum.
We used a maximum likelihood method (e.g.,
Crawford, Jauncey, \& Murdoch 1970) to fit the differential luminosity
functions of point sources (excluding nucleus, GCs, and stars) and
X-ray GCs with a broken power law. For X-ray point sources, the
luminosity break is at $9^{+2.0}_{-1.6}\times10^{37}$\lum, while
$\Gamma_1=0.74^{+0.84}_{-1.04}$, $\Gamma_2=2.58^{+0.35}_{-0.28}$, and $A=45.85$. For
X-ray GCs, the shape is nearly a cutoff power law with a luminosity
break at $3.1^{+0.26}_{-0.24}\times10^{38}$\lum,
$\Gamma_1=0.67^{+0.36}_{-0.39}$, $\Gamma_2=8.6^{+4.6}_{-3.2}$, and $A=5.90$.

\begin{inlinefigure}
\psfig{file=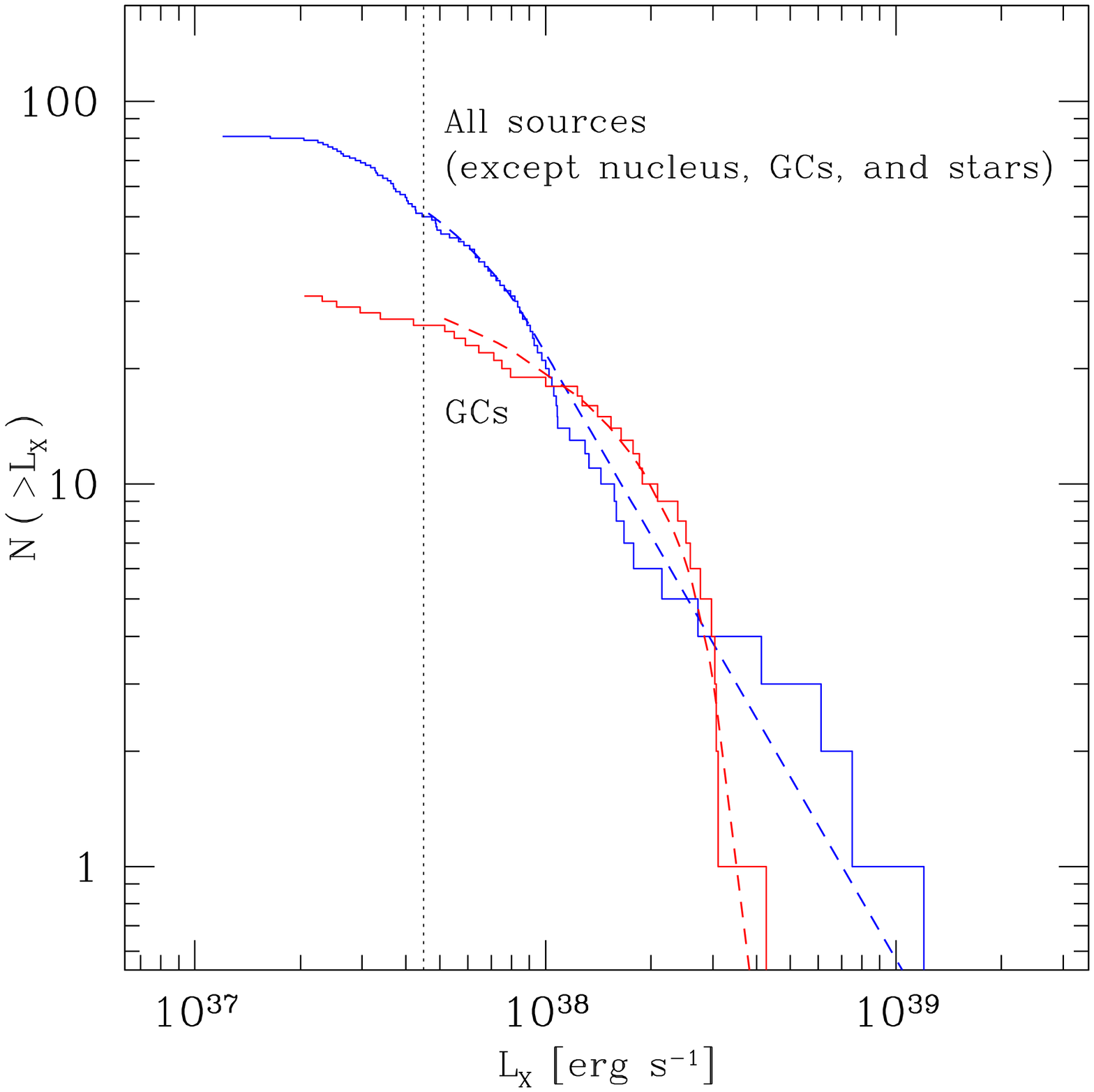,height=3in}
\caption{Cumulative luminosity functions and their best-fit model of
X-ray point sources (blue; excluding the nucleus 
and stars) and GCs (red) in M104. The vertical dotted line represents the
completeness limit ($4.5\times10^{37}$\lum) of our data. 
}
\end{inlinefigure}

\subsection{Color-color Diagram}

Many of the detected sources provide fewer than $100$ 
counts, which makes it difficult
to derive spectral parameters. However, hardness ratios can give a crude  
indication of the X-ray spectra. We therefore computed the hardness
ratios for each detected sources, for which we define HR1=(medium-soft)/(medium+soft) and
HR2=(hard-soft)/(hard+soft) to plot the color-color diagram as shown
in Figure 4. We calculated the $1\sigma$ uncertainties of the
hardness ratios by using a maximum likelihood method as used in the
{\it Einstein} catalog (Harris et al. 1993).

\section{Supersoft X-ray Sources}

SSSs have luminosities in the range $10^{35}-10^{39}$ erg s$^{-1}$.
and $k\, T$ in the range $10-100$ eV.
SSSs in
our Galaxy
 are largely obscured from us by
the interstellar medium (ISM); 
it is therefore possible that this poorly understood class
is in fact the dominant class of high luminosity X-ray binaries. 
Di Stefano \& Rappaport (1994) estimated, based on {\it ROSAT} observations and
simple models of galaxy absorption, that the Milky Way and M31 may each
harbor on the order of $1000$ SSSs with $L_X > 10^{37}$ erg s$^{-1}$
and $k\, T > 30$ eV. 
SSSs may significantly influence
their galactic environments, both as
efficient ionizers of the interstellar medium and possibly also as
progenitors of Type Ia supernovae. 

\begin{inlinefigure}
\psfig{file=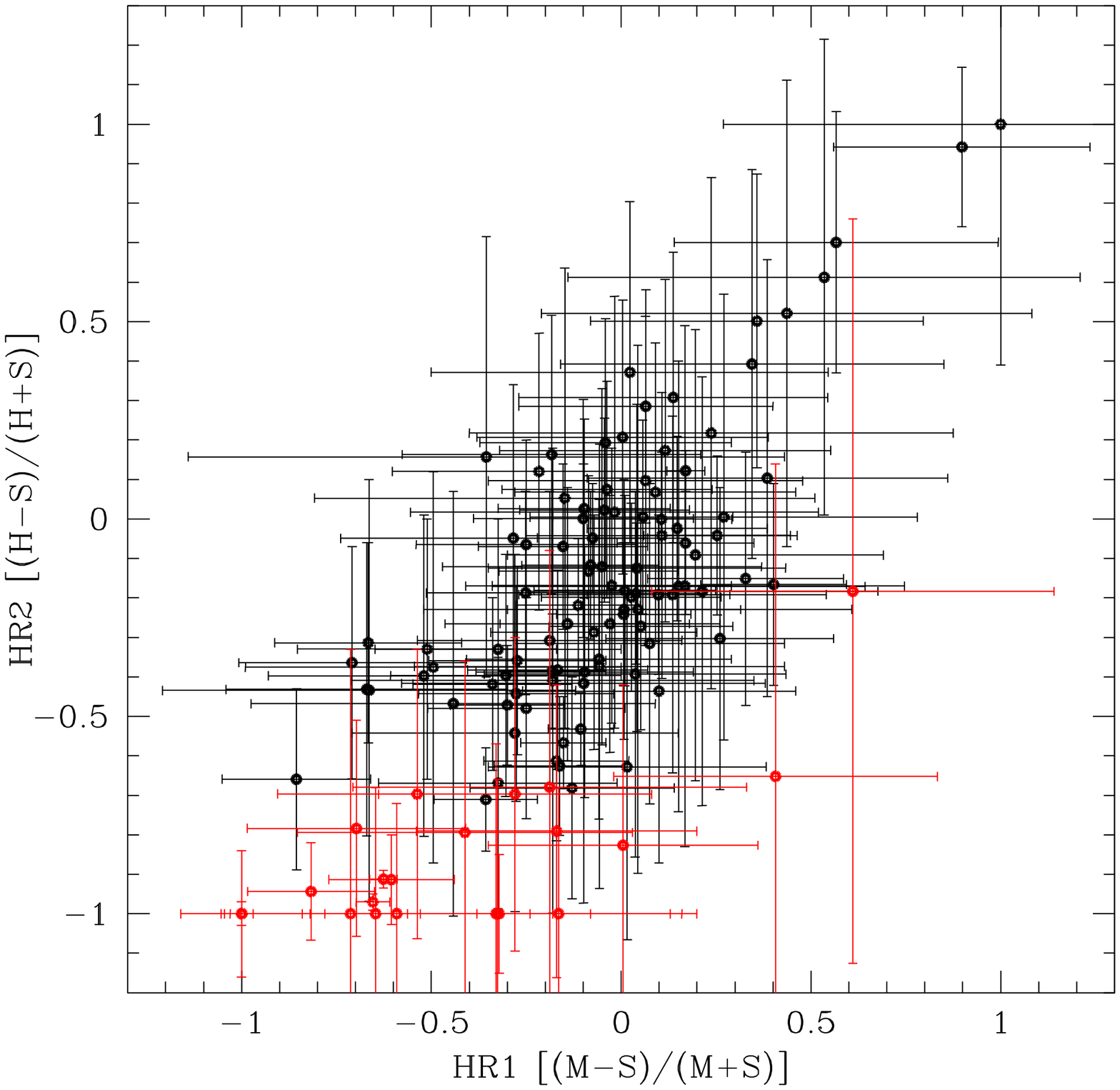,height=3.5in}
\caption{Color-color diagram of all X-ray point sources in M104. SSSs
are marked with red lines.} 
\end{inlinefigure}

Some SSSs are simply hot WDs (e.g., post novae),
pre-white-dwarfs (in planetary nebulae), or symbiotics.
However, the physical
nature of a majority of the SSSs with optical IDs
is not yet understood. These more mysterious sources include
CAL 83 and CAL 87 (Long, Helfand, \& Grabelsky
1981), as well
as $7$ sources discovered with {\it ROSAT} (see Greiner 2000 for details).
Black hole (BH; Cowley et al. 1990) and
accreting-neutron-star models (Greiner \etal\ 1991, Kylafis \& Xilouris 1993)
may apply to some SSSs.
Binary models which predict quasi-steady nuclear burning
of accreted matter on the surface of a WD
are considered promising (see, e.g., van den Heuvel et al. 1992.)
Nevertheless, studies of external galaxies
find some SSSs that may be BHs with $M\sim 100\, M_\odot$ (\rd\ \& Kong 2003a,
Kong \& \rd\ 2003).

The primary goals we hope to achieve by conducting SSS searches in
external galaxies are to gain an understanding of the physical nature or
natures of the sources and to eventually improve estimates of 
the sizes of galaxy populations of SSSs. 
We gain important insight by being able to
study the locations of SSSs relative to large-scale galaxy structures
(bulge, disk, halo), to study the stellar populations within which SSSs
are embedded, to establish luminosity distributions, and
spectral characteristics for large samples of sources, and to estimate
total populations as a function of galaxy properties. 
A challenge that must be
met by any such program of study is 
that the small number of counts we receive from X-ray sources 
in external galaxies makes spectral classification
difficult. Selection is further complicated by the
fact that we have only a small
sample of local (Galaxy and Magellanic Cloud) sources,
whose properties may not span the full range of SSS properties.
  \rd\ \& Kong (2003a, b) have developed an algorithm 
to select SSSs in external galaxies. Its application has
led to some intriguing results. 

In M101, SSSs appear to be primarily
associated with the spiral arms; this, combined with
the locations of several SSSs near HII regions and/or
OB associations, suggests that some SSSs are relatively young systems.
 In M31 (but not in M101), there is a bright bulge population
of SSSs, including  one source within a few parsecs of the nucleus.  
This suggests that some SSSs within $\sim 1$ kpc of a massive BH
may be the stripped cores of giants that have been tidally disrupted by the
BH (\rd\ et al. 2001). Several galaxies house ultraluminous SSSs 
(with $L_X > 10^{39}$ erg s$^{-1}$) that may be accreting 
intermediate-mass ($\sim 50-10^4 M_\odot$) BHs.  
These BH spectra typically include some photons with energies
above $1.1$ keV, even though $90\%$ of the flux
is carried by photons of lower energy.     
Elliptical galaxies (e.g., NGC 4697 (Sarazin et al. 2001)) 
contain SSSs, including SSSs in GCs (NGC 4472 (Friedman et al.2002)). 

We applied our algorithm 
to the X-ray sources we identified in M104.
In Table 1, SSSs marked either ``HR'' or 
``$3-\sigma$'' are likely to have spectra similar to those
of the SSSs studied so far the the Magellanic Clouds and Milky Way.
Sources labeled ``med'' or ``noh'', may be highly
absorbed, or may have somewhat higher 
temperatures, 
but $k\, T$ is still smaller than approximately $250$ eV.
Sources marked ``$\sigma$'', ``$3\sigma\, 1$'' or ``HR1'' could
have temperatures in the same general range as ``med'' or
``noh'' SSSs; our studies of SSSs in other galaxies
show, however, that some have 
spectra as soft as the  ``HR'' or 
``$3\sigma$'' sources, but also include a small ($< 10\%$) harder component.

Roughly one in six X-ray sources in M104 satisfies the conditions designed
to identify SSSs. 
Below we discuss the luminosity distribution of the SSSs detected
in M104, and an intermediate-mass BH model for the brightest SSS in M104.
The unique feature of M104, is that one can identify and
study the disk, bulge,
and halo populations of SSSs; these are discussed in \S 4.3-\S 4.5.

\subsection{Luminosity Distribution of SSSs} 

The dimmest GC X-ray source ($2 \times 10^{37}$erg s$^{-1}$)
 and the $6$ dimmest non-GC X-ray sources (starting at
$1.4 \times 10^{37}$ erg s$^{-1}$) are all SSSs.
In fact, of $82$ non-GC X-ray sources, $15$ of the $30$ least luminous, are SSSs;
these represent $3/4$ of the non-stellar SSSs. The dim SSSs
each  
provide $8-20$ photons. 

We expect low-count-rate SSSs for two reasons.
First, the intrinsic SSS luminosity function of some SSSs is dominated by
low-L sources. The luminosity of nuclear-burning WDs, e.g.,
increases with increasing WD mass, with $L$ nearing the 
Eddington limit and $k\, T$ perhaps in excess of $100$ eV,
 as the WD mass
approaches the Chandrasekhar mass. Since, however, 
WDs with lower mass are more common, most nuclear-burning WDs
should be less luminous. In addition, Greiner et al. (1999) 
and Greiner \& \rd\ 1999, established that there  is a lower-luminosity
extension of the class of SSSs, which presumably corresponds to 
nuclear-burning WDs of even lower mass than those in the systems first
observed with {\it Einstein} and {\it ROSAT}.   
The number of systems
in our Galaxy capable of appearing as such lower-luminosity SSSs
 may be approximately $10,000.$ 
Second, the effects of absorption on the radiation emitted by SSSs
is severe. The $18.5$ ksec observation of M104 was likely to only     
detect those SSSs on the outer edge of the
 side of the disk and bulge nearest to us.

In spite of the numerical dominance of the low count rate SSSs,
SSSs are also well represented among the most luminous SSSs.
Three of the $12$ brightest M104 X-ray sources are SSSs--these include 
one of the
brightest non-nuclear sources, X82.

\subsection{Intermediate-Mass BH Model for the Brightest SSS} 

X82 is an SSS located in the bulge of M104.
The spectral fit is shown in Figure 3. The spectral parameters
(Table 2) are: $k\, T = 180$ eV, $L_X = 8.9 \times 10^{38}$ erg s$^{-1}$,
$N_H = 0.9^{+4.6}_{-0.9} \times 10^{20}$ cm$^{-2}$. This fit cannot
provide a unique physical interpretation.  It is, however,
consistent with the model of an intermediate-mass BH.
Consider a geometrically thin but optically thick accretion disk, and
identify the inner edge of the disk with the last stable
orbit around an accreting BH. If we assume an emission efficiency
of $10\%$ (i.e., that $10\%$ of the rest energy of matter falling into the BH
is radiated by the disk), then

\begin{equation}   
M_{BH}=10^3\, M_\odot\ \Big[{{49\, eV}\over{k\, T}}\Big]^2\  
                \Big[{{L}\over{2.9 \times 10^{37} erg/s}}\Big]^{{1}\over{2}}  
\end{equation}   

For the spectral parameters derived in the fit, we find that the mass
of the BH would be $\sim 400\, M_\odot$. This is likely to be a lower limit,
since spectral hardening effects, orientation effects,
 and spin would all tend to increase
the derived value of the mass. 
 
\subsection{SSSs in the Bulge of M104}

There is a clear over density of SSSs in the region near the nucleus:
$4$ SSSs are located within $\sim 1$ kpc of the galaxy center, 
and an additional $3$ SSSs are located within $1.5$ kpc of the 
center. Since this places $1/3$ of the SSSs in roughly
$2\%$ of the area containing SSSs, it is clear that this 
overdensity implies that some of these $7$ SSSs
are physically close to the nucleus.
In M31 there are SSSs within several parsecs
of the nucleus, and members of the bulge population 
are among the most luminous 
SSSs in the galaxy. While it is possible
that all or most of these sources are simply descended
from the stellar populations that inhabit the bulge, the presence of  
SSS very close to the nucleus of M31 has suggested that some bulge
SSSs may be the stripped cores of stars that have been
 tidally disrupted by a central BH. If this is so, then
galaxies with massive BHs, such as M104, with an estimated
BH mass of $10^9 M_\odot$, should have a central overdensity of
SSS.

\subsection{SSSs in the Disk of M104}

Given the small numbers of SSSs observed near the disk, coupled
with the large size of the bulge, it is difficult to definitely identify SSSs
as members of a disk population. Indeed, SSSs in or near the disk
are the ones most likely to be subject to absorption.
The locations of X21 (9 counts), X37 (9 counts), X110 (15 counts), 
and possibly X102 (20 counts) are consistent with membership
in the disk. Indeed the distributions of counts in 
S, M, and H for at least $3$ of these sources (X37, X110, and X102)
would be consistent with what is expected for a highly luminous,
highly absorbed 
source with $k\, T \sim 100$ eV.

\subsection{SSSs in the Halo of M104}  
One of the most exciting aspects of the M104 observations is that  
for the first time, we can 
unambiguously identify SSSs located out of the disk
of a spiral galaxy. (See also \rd\ et al.\, 2003a, on M31.) 
The $2$ softest sources, X71, and X86, are each located
$> 4$ kpc away (north and south, respectively) 
from the galactic disk. Neither is associated with a known GC or with
any other optical counterpart on the DSS image. From neither
source have we received photons with energies $> 1.1$ keV.
The estimated luminosity of X71 is $1.4 \times 10^{37}$ erg s$^{-1}$,
while for X86 we derive $1.6 \times 10^{38}$ erg s$^{-1}$.
(These estimates use a model with $k\, T = 125$ eV and 
$N_H =  5 \times 10^{20}$ cm$^{-2}$.)     
If these are not foreground or background  
objects, they 
may be good candidates for the NBWD  model.
Since such systems should not have experienced prior 
supernova explosions, they are unlikely to have been ejected
from the disk. They are therefore likely descendants of
an old halo stellar population.   
A particularly interesting SSS is  X35, which is in a GC.
(From its position alone, it could be part of the bulge or halo.)
This source does emit photons between 1.1 and 2 keV. 
The source properties are consistent with those of a BH accretor
with mass equal to  
$\sim 50 M_\odot$ (See equation (2)). This is similar to the situation we have found in
NGC 4472, where at least $5$ SSSs are associated with GCs, several
amenable to a similar interpretation (Friedman et al. 2002,  
\rd\ et al.\, 2003b).   

\subsection{Relationship between SSSs and Planetary Nebulae}

In the region covered by the S3 CCD, $249$ PNe have been
identified. None of these is near enough to an X-ray source
to suggest a physical association. The fact that no PNe are 
associated with SSSs is interesting. Because SSSs
emit highly ionizing radiation. Indeed, one SSS in the Magellanic Clouds is
a PN (Wang 1991). 
SSSs that are binary systems can ionize the interstellar medium in which
they are embedded. Those located in regions in which the ISM has
a density greater than $\sim 1$ atom cm$^{-3}$,
can create circumstellar ionization nebulae 
with very large (radiation-limited) radii ($>10$ pc)
(Rappaport et al. 1994).
CAL 83, located in the Magellanic Clouds  is embedded
in such a nebula (Remillard, Rappaport, \& Macri 1995). 
With two of twelve Magellanic Cloud SSSs associated with nebulae,
and with $\sim 1000$ SSSs predicted in galaxies such as M31,
we might expect there to be dozens of such associations.
The nebulae could be and  would likely be included among PNe lists
by surveys using the methodology Ford et al (1996) used to
identify PNe in M104 (\rd , Paerels, \& Rappaport 1995).
Because  only a small fraction of all SSSs can be detected,
the number of observable PN/SSS may be only $1-10\%$ of the
number of physical associations. We might expect
some matches for M104 (\rd , Paerels, \& Rappaport 1995).
The lack
of any such associations may mean that SSSs are  
typically in regions that have a very low density ISM, or 
that they typically are active for periods smaller than
the 
recombination time (typically $\sim 10^5$ years), 
with long gaps between ``on'' times (Chiang \& Rappaport 1996).  
But, since it is not clear that either of these possible explanations 
should or does apply, it would be helpful to
push the completeness limit for SSSs to lower X-ray luminosities.
If the link between nebulae and SSSs can be established and if, as
predicted (Rappaport et al. 1994, \rd\ et al 1995), SSS nebulae
can be distinguished from PNe by a careful examination of the
full spectrum of excitation lines, then the nebulae may 
provide an independent
guide to the size of SSS populations in external galaxies.  

\subsection{Relationship between SSSs and Supernova Remnants}

The very soft X-ray emission of SSSs can 
resemble X-ray supernova remnants (SNRs). 
There are, e.g., two SSSs in M31 that 
were subsequently identified as SNRs based on 
extended morphology and multi-wavelength observations 
(\rd\ et al. 2003, Kong et al. 2002b, 2003b). 
These two SNRs appear to be constant X-ray sources, 
which is consistent with the expected picture of SNRs.  
On the other hand, most of the SSSs in external galaxies are 
X-ray variables (Di\,Stefano \& Kong 2003a; Kong \& Di\,Stefano 2003).
Although it is extremely difficult to study the morphology of X-ray sources at the distance of M104, and high resolution narrow-band optical imaging is not available, we can study the time variability of these X-ray sources to investigate their nature..

\section{GC X-Ray Sources}

\subsection{Significance}

Galactic GCs house approximately $100$ times as many X-ray sources
as would be expected based on their mass, suggesting that
the efficiency of forming X-ray sources must be higher in GCs than
elsewhere in the Galaxy (Clack 1975).
A possible explanation
is that the high stellar density near the cluster centers
(as large as $10^6$ stars pc$^{-3}$), can lead to stellar interactions
in which close binaries containing white dwarfs (WDs) and neutron stars (NSs)
can be formed. Eventually, either because dissipative processes (gravitational
radiation or magnetic braking) bring the companion stars close enough for
contact to be established, or else because the non-compact star
begins to evolve and comes to fill its Roche lobe, mass transfer begins.
Because the mass of the donor is generally limited by the turn off mass
($\sim 0.8\, M_\odot,$) these systems are classified as low-mass
X-ray binaries (LMXBs). 
Detailed calculations of interactions that form close binaries,
follow the subsequent binary evolution, and include interactions which
may occur after the close binaries are formed,
show that a combination of $2-,$ $3-,$ and $4-$body
interactions can create X-ray binaries and lead them to be
ejected
to the outer parts of the cluster. 
(See, e.g., Phinney \& Sigurdsson 1991.) 
It is considered likely 
that close binaries can be ejected from GCs, since the amount
of energy needed to escape the cluster
is a small fraction of the binary's binding energy.  

LMXBs can also be formed in the field, because a subset
of all primordial binaries are expected to evolve into close binaries
in which a neutron star accretes mass from a low-mass companion. 
Before a primordial binary can turn on as an X-ray binary, 
several processes must occur. The formation of the neutron star
occurs quickly, but the spiral in of the companion, or the companion's
stellar evolution, can require several billion years. 
Because there are many uncertainties in the evolution of
these systems and in the evolution of galaxy populations,
it is difficult to accurately predict the numbers of LMXBs
expected per unit mass, due to a single burst of early star formation. 
It has been  suggested, however, that such systems 
may not be able to produce enough LMXBs to explain the numbers currently being
discovered in early-type galaxies
White, Sarazin, \& Kulkarni(2002). (It is important to note that
some of the early-type
galaxies we observe may have had more complicated histories,
with multiple epochs of star formation.)   
If not, then it is possible that the LMXBs we do see
are systems that have been ejected from GCs.
Indeed, testing this hypothesis has been an important motivator
for study of the GC/LMXB connection.
Meaningful tests must include the study of a set of galaxies,
including spirals, to search for evidence of
differences in the LMXB/GC connection
 as exhibited by early- and late-type galaxies.
Until now, M31 was the only spiral with bright enough
GC X-ray sources to enable a comparison with the X-ray source population
in distant ellipticals. 
M104 is therefore an important addition to the set of 
galaxies in which we can hope to study the LMXB/GC connection.

Below we focus our discussion on the GCs studied from the ground  
which have been matched with X-ray sources. 

\subsection{Optical Selection of Globular Clusters}

        Deep, wide-field images of M104 have been acquired with
the CFHT12K mosaic camera at the Canada-France-Hawaii Telescope
(CFHT) in order to investigate the optical characteristics of
its large globular cluster system (GCS).  A complete description
of this photometry will appear in VanDalfsen et al. (2003).
In summary, the GCS exhibits a classically bimodal color distribution,
with ``blue'' (metal-poor) and ``red'' (metal-rich) subpopulations
centered at corresponding metallicities of [Fe/H](blue) $= -1.1$ and 
[Fe/H](red) $= -0.1$.  The redder population -- which makes up about
one-third of the total GCS -- is more centrally concentrated in spatial
distribution than the bluer clusters.  

In terms of the luminosity limit of our
optical survey, our data reach nearly two magnitudes fainter than
the normal ``turnover'' point (peak frequency) of the globular
cluster luminosity function (GCLF). The GCLF turnover is at
an absolute visual magnitude $M_V = -7.6$, corresponding to a
cluster mass of $M \simeq 1.3 \times 10^5 M_{\odot}$, with little difference
between the blue and red clusters.  In short, the characteristics
of the GCS strongly resemble those of the ``standard'' Milky Way
GCS, simply scaled up to a total cluster population 7 or 8 times larger.

\subsection{Properties of Matches}

The procedure we have used to match GCs with X-ray sources is described in \S 3.  
We find $25$ matches.

\noindent{$\bullet$}
  Of these, 19 objects lie in the
  metal-rich side, while 6 objects lie in the metal-poor side,
  giving a 1:3 ratio of blue:red objects.
  However, a random sampling of globulars would globally yield
  a 2:1 ratio of blue:red objects, and approaching a 1:1 ratio
  of blue:red very close to M104.
  There also seems to be a lack of matches which are both bright
  and blue. We therefore find a result similar to others reported for
NGC 4472 (Maccarone et al 2003, Kundu et al. 2002) and for the 
Milky Way and M31 (Bellazzini et al. 1995, but see also \rd\ et al. 2002),
that X-ray sources are preferentially found in GCs that are metal rich.
(See Figure 6) 

\noindent{$\bullet$}
Figure 7 illustrates the relationship between X-ray
luminosity and the magnitude and color of the clusters
housing X-ray sources. Two new features are worthy of note.
First, the most luminous clusters in both $B$ and $R$
that have X-ray sources
house only highly luminous X-ray sources ($L_X > 10^{38}$
erg s$^{-1}$). (In contrast, 
some clusters which are not as optically bright, 
house X-ray sources of lower luminosity, while others
 harbor bright X-ray sources.) 

\noindent{$\bullet$}
Second, for X-ray sources with $L$ greater than approximately
$2.0 \times 10^{38}$ erg s$^{-1}$, there is no obvious preference
for GCs that are optically redder.  
 
\noindent{$\bullet$}
None of the GC X-ray sources have luminosities greater than
the Eddington limit for a $1.4 M_\odot$ object Given the small numbers
of sources at the highest luminosities, 
and the fact that transient behavior is common in LMXBs, 
this situation could be 
different in future observations.

\noindent{$\bullet$}
Although the $4$ brightest X-ray sources are not in GCs, GC X-ray sources dominate
(in fact represent $\sim 68\%$ of all sources in the luminosity range 
$\sim 1.2-4 \times 10^{38}$ erg s$^{-1}$).
That is, bright X-ray sources are most likely to be found in GC. 
 
\noindent{$\bullet$}
As described in \S 4, 
one GC houses a soft X-ray source, X35, that could be an accreting BH
with a mass of approximately $50\, M_\odot$.

\subsection{Interpretation}

The 4 brightest X-ray sources in M104 are super-Eddington for a
$1.4 M_\odot$ object; these 4 sources are not located in GCs.
Nevertheless, the majority of (68\%) all X-ray sources with 
$1.2 \times 10^{38}$erg s$^{-1}<L_x<4 \times 10^{38}$erg s$^{-1}$
are in GCs. Were M104 located far enough from us, the majority 
of sources we observe could be in GCs. 
M104 provides a second example of a spiral galaxy
(M31 being the first) in which the majority of the brightest sources
are located in GCs. 

Our data are consistent with previous findings that
X-ray sources are found preferentially in red, hence metal-rich, GCs.
An interesting twist, however, is that this preference
for red GCs, is not expressed by the brightest X-ray sources.
The least luminous GC X-ray sources in our sample do, on the other hand,
have a strong preference for red GCs.

\begin{inlinefigure}
\psfig{file=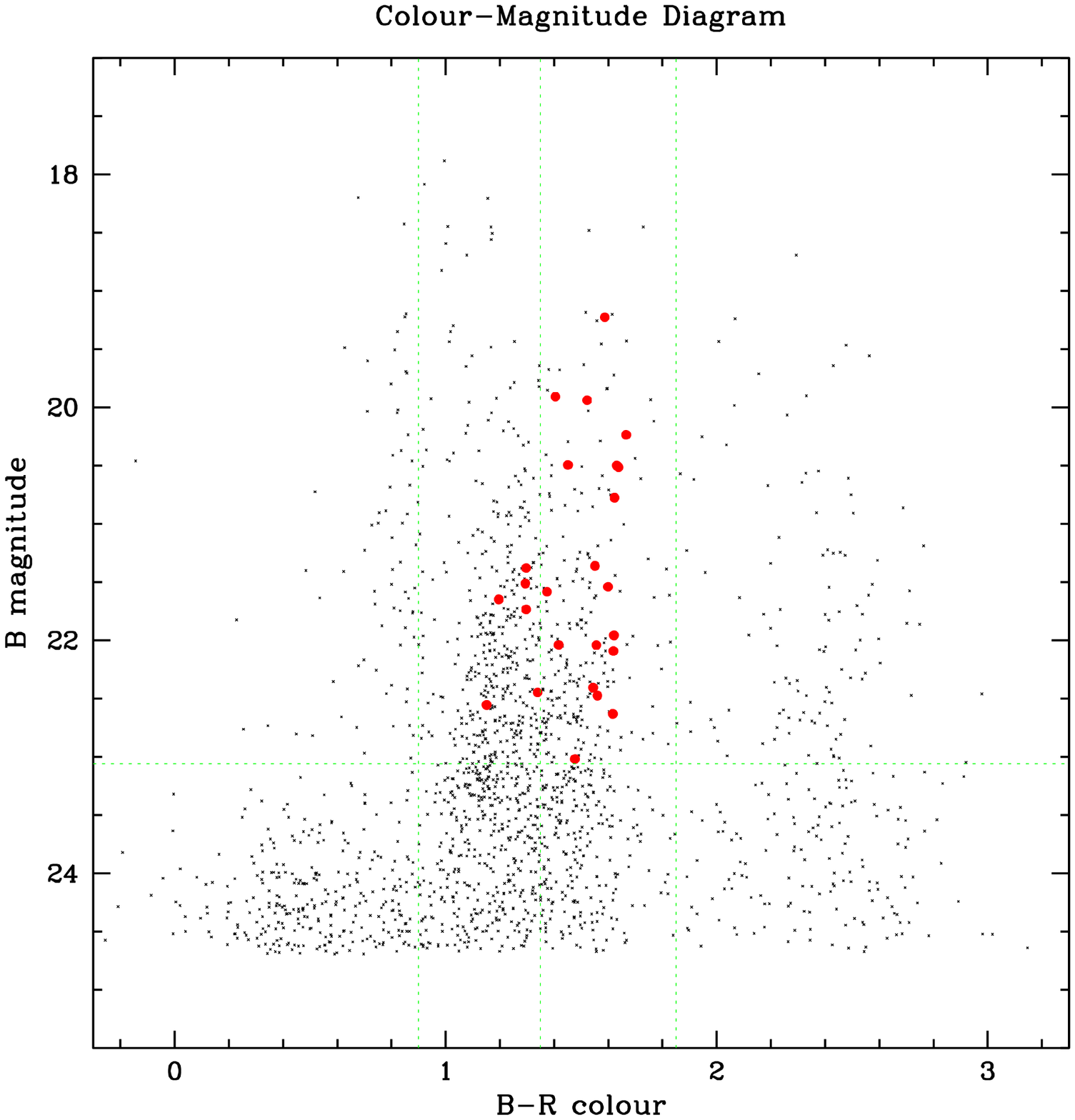,height=3in}
\caption{Color-magnitude diagram of all of the objects detected by the
CFHT survey located 
  within about 9 arcminutes (radius) of M104.  This diagram shows 
  globular clusters and non-cluster contaminants.  The 25 X-ray matches
  are shown by large circles.  The horizontal line (at $B=23.06$) is 
  the location of the GCLF turnover magnitude.  The left and right
  vertical lines are the color limits used to identify GCs; they  
  extend roughly $3-\sigma$ out from
  the GCS population, covering $(B-R)$ from $0.90 - 1.85.$  The middle 
  vertical line represents the division between the blue (metal-poor)
and red (metal-rich) populations, at $(B-R)=1.35.$}
\end{inlinefigure}

\begin{inlinefigure}
\psfig{file=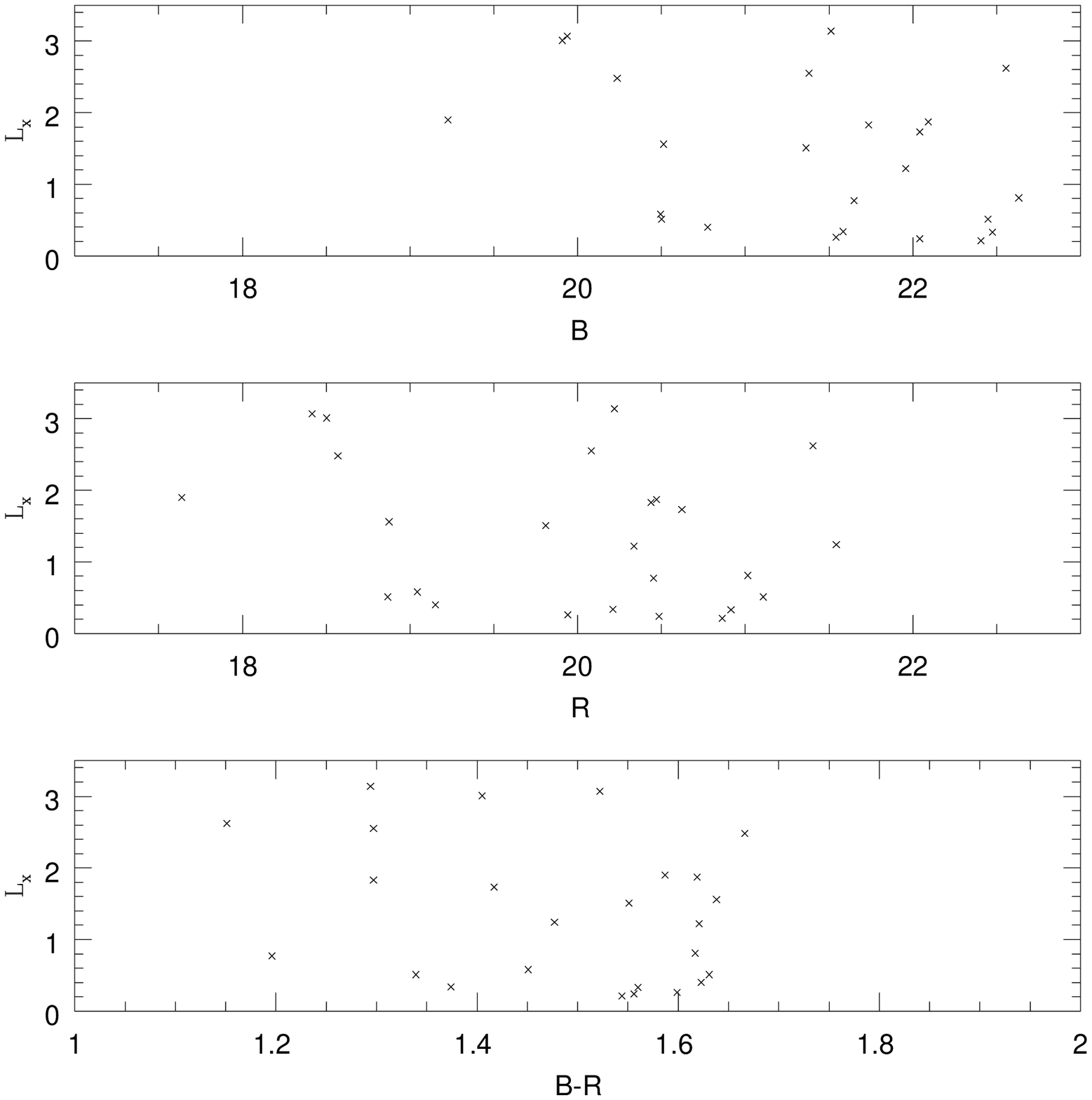,height=3in}
\caption{
The X-ray luminosity, in units of $10^{38}$ erg s$^{-1}$, is plotted
versus: $B$ magnitude (top panel); $R$ magnitude (middle panel);
$B-R$ (bottom panel). Each point represent a single association
between an X-ray source with X-ray luminosity $L_X$ and an CFHT-detected
 GC with
optical properties described by $B,$ $R,$ or $B-V.$   
}
\end{inlinefigure}

\subsection{Models}

Below we discuss possible explanations for the trends 
apparent in the data.
 
\subsubsection{Thermal Time Scale Mass Transfer}

The properties of the luminous X-ray source in the M31 GC Bo 375,
suggested a new type of GC binary evolution model (\rd\ et al. 2002).
In this model, mass transfer is driven by the thermal
time scale readjustment of the donor star to changes in 
the size of its Roche lobe. 
Mass transfer typically starts when the 
donor star has a mass that is within a few tenths of 
a solar mass of the NS, and is 
also
 slightly evolved (with core mass $\sim 0.1-0.25$).
Because the donor is more massive than $0.8\, M_\odot,$
this type of system can be found only in younger GCs and/or
in GCs with such a high rate of stellar interactions that
at least one NS can   
be expected to have a blue straggler donor.
One proposed test of this model was to check if bright X-ray sources
in the GCSs of external galaxies are likely to be found in young and/or
optically bright GCs. This is exactly what we find in the
Sombrero. Thus, although the model is not proved, M104
does provide some indirect evidence in its favor.

\subsubsection{Multiplicity and Ejection}

The correlation between optically bright GCs and high-luminosity
X-ray sources is intriguing.  It is likely that the large optical
flux is indicative of a massive GC, in which stellar interactions
producing X-ray binaries may be more frequent. More frequent
interactions may lead to more X-ray sources per cluster.
There are two possible checks for multiplicity.  
 
\noindent{\sl Time Variability:\ }
GCs that do not house multiples, or that include only a small number
($\sim 2-3$)  of individual components, could exhibit
significant 
X-ray
variability 
(by factors of $2-3$) 
over the same time scales on which X-ray binaries
are typically variable. Thus, over time scales of days, weeks,   
or months, we could expect to find evidence of variability.
In contrast, GCs 
that have a high level of X-ray source multiplicity--i.e., in which the
total flux is a composite consisting of comparable contributions from
more than a few sources--should be almost constant, since it is unlikely
that many of the independent components will have $L$ declining
or increasing simultaneously.

\noindent{\sl Angular Resolvability:\ }
At the distance of M104, one arcsecond corresponds to
roughly $43$ pc, larger than the radii of typical GCs.
Most of the GC X-ray sources in our sample are located away from
the aim point, in regions where the angular resolution may be 
as large as several arcseconds. If, therefore, some of the 
high luminosity sources we detect in bright GCs are actually
composites, one or more of the individual components may consist
of X-ray sources that have been ejected from the cluster.
It is possible that a subset of such systems  could be resolved by
pointed observations with {\it Chandra's} High Resolution Camera.

\subsubsection{Giant Donors}

The observations we have described introduce a
puzzle: 
the brightest blue GCs do not host X-ray sources,
while dimmer blue GCs can house X-ray sources
that are among the brightest in the galaxy.
%Thus, although the blue clusters with X-ray sources in our sample 
%all are brighter than the GCLF turnover, 
%and thus have masses between $10^5$ and $10^6 M_\odot,$
We therefore seek to understand how they come to house
extremely bright X-ray sources,
while 
the brightest blue GCs do not appear to harbor X-ray sources,
even X-ray sources that are $\sim 10$ times dimmer.
 
The thermal-time-scale scenario described above
is likely to work in either young GCs, where the turn-off
mass may be larger than $0.8\, M_\odot$, or in
very massive GCs, in which the probability of a NS orbiting
a blue straggler donor is larger. Similarly, multiplicity
would seem to be most likely in large GCs with a high
level of concentration and hence, high central surface brightness.

The problem posed above would seem to require another solution.
Presumably the donor stars are old, and are less massive than
their NS companions. In such cases, the mass transfer rate
can be high if the donor is a giant. Such systems will exist
in a GC only if a binary with total mass $\sim 2.2\, M_\odot$,
and a separation of $\sim 50-1000\, R_\odot$
can survive. Some such binaries are
considered ``soft'' in the GC environment, having
less binding energy than the ambient kinetic energy of
surrounding stars. Soft binaries have a good chance of being
disrupted by a combination of distant
interactions acting over time. Others of these binaries may
not be formally ``soft'', but, since they are likely to
be found near the cluster center, where close interactions are most
frequent; they can also be disrupted. Both mechanisms are
least efficient at destroying the binaries with
large separations in less massive and less highly concentrated
GCs. It may therefore be that the most
X-ray luminous X-ray sources in the less bright blue GCs  
are accreting NSs with giant donors.

It would be difficult to obtain direct evidence that a
given GC X-ray source, even in M31, was an example of a NS
accreting mass from a giant donor.
A useful form of indirect evidence could, however, come from
GCs in our own Galaxy. 
Specifically, we could attempt to establish the existence, in 
some clusters with low present-day interaction probabilities,
of relatively luminous 
($L > 10^{32}$ erg s$^{-1}$) CVs with giant donors. 
Galactic GCs are near enough that the optical properties of the 
X-ray binaries can be studied with {\it HST}, allowing
general features of the model to be checked.
WDs accreting from giant donors, which cannot presently 
be studied in external galaxies, 
may be more common analogs of the accreting NSs with giant donors.
As such, they
could provide insight into the environments that nurture such
systems.

\section{Conclusion and Prospects}

M104 may be one of the richest theatres in which to study
X-ray emission from galaxies. In this paper we have focused on the
emission from non-nuclear point sources, 
since the diffuse emission (see Figure 2)
which motivated this observation is the subject of a separate paper
(Forman et al. 2003; see also Delain et al.\, 2001), 
as is the nuclear source (Pellegrini et al.\, 2003).

The primary science results we have extracted so far from the $18.5$ ksec
observation  revolve around SSSs and the connection between GCs
and LMXBs.   
We find SSSs inhabiting the bulge, disk and halo
of M104. At least $4$ of the halo sources are very soft,
likely as soft as the SSSs known in the Magellanic Clouds.
These are good 
candidates for nuclear-burning accreting WDs.  
The discovery of these sources several kiloparsecs
from the disk, indicates that they are old systems,
as should be expected for a subset of nuclear-burning WDs.
In the halo as well as in the disk and bulge, we also
find SSSs from which we have collected a small number of photons with
energies above $1.1$ keV. In some cases, photon
statistics or the 
incomplete subtraction of background due to diffuse emission
could be responsible for the harder photons. In most cases,
however, the photons are likely to emanate from the system
itself. Some SSSs exhibiting this harder emission 
may be supernova remnants. Since, however, similar sources have
been discovered in other galaxies, including M31, M101, NGC 4472,
M51, and M83, and since we have discovered that many such SSSs
are variable on scales of months to years, it is likely that a large fraction
are X-ray binaries. 
The harder photons from some SSS binaries
may support a model in which the accretor is an intermediate-mass 
BH.   
We find a significant overdensity of SSSs within $1$ kpc 
of the nucleus; future work will assess whether the overdensity
is more pronounced for SSSs than for other X-ray sources.
If it is, this could be indicative of tidal disruptions  
of giants by the $\sim 10^9 M_\odot$ BH at the center of
the Sombrero. 

Regarding the LMXB/GC connection,
we find that the fraction of bright X-ray sources in GCs is similar
across galaxy types. 
We also find that those optically bright GCs 
that have X-ray sources house only bright X-ray sources.
M104 provides another example of a galaxy in which
X-ray sources are most likely to be found in red GCs; these clusters
are presumably more metal rich and younger than bluer clusters.
Interestingly enough, this preference for redder clusters does
not seem to be expressed by the brightest ($L_X > 10^{38}$ erg s$^{-1}$)
X-ray sources. Another interesting fact is that 
M104's optically brightest bluer clusters do
not seem to house X-ray sources. 
We have sketched several models which can  
help to explain the trends seen in the data.
Thermal-time-scale mass transfer can, in younger or
more massive clusters, produce high 
luminosity systems. Multiplicity can play a 
role in some massive clusters. Accretion by a NS from a giant
could be occurring in some blue clusters which house 
X-ray bright sources.  
 
\subsection{Prospects}

An ultimate goal of X-ray observations of galaxies is to
better understand the natures of the point sources and the 
processes by which stellar populations create them. To that end,
the fact that M104 has a large X-ray source population, and is also
well studied at other wavelengths, supports
the process of relating X-ray sources to their environments. Its 
nearly edge-on orientation
is also a boon, because it provides a rare opportunity
to explore regions out of the disk of a spiral galaxy
with a minimum of projection effects.

The Sombrero is not only an important 
galaxy to study in its own right,  
but it is also an important member of a set of external galaxies
whose study will allow us to probe environmental effects
on the creation, evolution, and properties of X-ray sources.
Together with well-studied galaxies such as the Milky Way, M101, M83, and M81,
it provides an example of an X-ray active spiral galaxy, supplementing information
derived from the others because it is a bulge-dominated spiral
and also because of its edge-on orientation.
Consider the contrast with, e.g., M101, which is located at a comparable
distance, but which is viewed nearly face-on.
In M101  
X-ray sources can be directly identified with disk stellar
populations, but
 it is not
possible to know which X-ray sources are actually located away from the
galactic plane. The combination of studies of M101 and
M104
can therefore
play complementary roles. Another key difference is that
the central region of M101 is not known to house a
central BH, while M104 harbors a central BH with $10^9 M_\odot.$
Already we know that the central concentration of X-ray sources
in M101 is not as extreme as in M104, and that there are no bright
SSSs with $1$ kpc of the nucleus (\rd\ \& Kong 2003a, b). 
M31 does have a massive BH, but it is $30$ times less massive than the
Sombrero's; it will be important to compare the densities and properties 
of X-ray sources among these galaxies. 
Because the Sombrero provides a link between disk-dominated and 
bulge-dominated galaxies, it will  
also be productive
to compare the X-ray source population of M104 with those of elliptical
galaxies, such as NGC 4472 (Kundu et al. 2002; Maccarone et
al. 2003), NGC 4697 (Sarazin et al. 2001), and NGC 1399 (Angelini et al. 2001).

\subsubsection{Future Observations}

Future observations can not only 
provide a profile of the X-ray source distribution in the Sombrero,
but can answer a range of science questions
related to variability and to the environmental influences
that affect the formation and evolution of X-ray sources in galaxies.

\begin{acknowledgements}
This work was supported by NASA under an LTSA grant,
NAG5-10705. A.K.H.K. acknowledges support from the Croucher Foundation.
W.E.H.\, is pleased to acknowledge financial support from the Natural
Sciences and Engineering Research Council of Canada.
R.D.\, would like to thank W. Forman and C. Jones for their careful 
reading of the paper and for comments, and R. Kraft and F.A. Primini for
interesting conversations. We thank T.A. Russo for her careful reading
of the manuscript. 
\end{acknowledgements}

\begin{deluxetable}{lccccccccc}
\tabletypesize{\small}
\tablecaption{Source list}
\tablewidth{0pt}
%{\centering
%\footnotesize
%\begin{tabular}{lcccccccc}
%\hline
%\hline
\tablehead{\multicolumn{1}{c}{Object}& R.A.& Dec. & \multicolumn{3}{c}{Net Counts} & $L_X$\tablenotemark{a}
& Optical Magnitude  & Note\\
%\tablehead{Object& R.A.& Dec. & &Net Counts& & $L_X$\tablenotemark{a}
%& Optical Magnitude  & Note\\
 & (h:m:s)& $(^{\circ}:\arcmin:\arcsec)$ & Soft & Medium & Hard &
($\times 10^{38}$) & & &}
%\hline
\startdata
X1& 12:39:38.7&-11:38:51.1&$21.7\pm4.8$&$11.1\pm3.4$&$-0.5\pm0.2$&0.69&$B=13.8$,$R=12.5$,$K=10.9$&star,SSS-med\\
X2& 12:39:42.3&-11:36:56.1&$7.6\pm2.8$&$9.4\pm3.0$&$6.9\pm2.6$& 0.73\\
X3& 12:39:43.1&-11:36:44.2&$2.9\pm1.8$&$6.0\pm2.5$&$6.7\pm2.7$& 0.48\\
X4& 12:39:44.2&-11:36:00.5&$30.3\pm5.6$&$21.4\pm4.6$&$7.2\pm2.8$&1.76 \\
X5& 12:39:45.2&-11:38:49.5&$630.2\pm25.1$&$144.9\pm12.0$&$28.8\pm5.3$&19.20 &$B=10.3$,$R=9.7$,$K=8.4$&star,SSS-$\sigma$\\
X6& 12:39:45.2&-11:36:00.4&$36.6\pm6.1$&$39.5\pm6.3$&$25.0\pm5.1$& 2.48&$B=20.0$,$R=18.5$,$K=15.6$&CFHT\\
X7& 12:39:45.6&-11:39:33.1&$7.2\pm2.6$&$4.9\pm2.2$&$10.0\pm3.1$&0.67 &$B=19.9$,$R=18.9$&star\\
X8& 12:39:47.2&-11:38:14.6&$9.3\pm3.0$&$6.3\pm2.5$&$4.9\pm2.3$& 0.63\\
X9& 12:39:48.6&-11:37:13.0&$87.5\pm9.3$&$64.3\pm8.0$&$24.2\pm4.9$& 5.42\\
X10& 12:39:50.1&-11:37:41.7&$2.8\pm1.8$&$3.1\pm1.8$&$1.8\pm1.4$& 0.24\\
X11& 12:39:50.5&-11:39:14.5&$19.1\pm4.4$&$19.5\pm4.4$&$13.2\pm3.6$&1.56 &$B=20.2$,$R=19.0$&star\\
X12& 12:39:50.9&-11:38:23.3&$31.5\pm5.6$&$44.4\pm6.6$&$27.9\pm5.3$& 3.14&&CFHT\\
X13& 12:39:51.1&-11:36:47.7&$7.6\pm2.9$&$9.3\pm3.0$&$3.0\pm1.7$& 0.59\\
X14& 12:39:51.1&-11:39:28.4&$-0.6\pm0.2$&$3.9\pm2.0$&$4.9\pm2.2$&0.26 \\
X15& 12:39:51.1&-11:38:33.7&$9.3\pm3.0$&$5.2\pm2.2$&$1.6\pm1.4$&0.40 &&SSS-$3\sigma1$\\
X16& 12:39:51.9&-11:35:51.9&$38.2\pm6.2$&$27.5\pm5.3$&$8.8\pm2.9$&2.29 &&CFHT\\
X17& 12:39:53.4&-11:37:09.3&$6.1\pm2.7$&$8.0\pm2.8$&$4.1\pm2.0$&0.57 \\
X18& 12:39:54.0&-11:38:25.1&$30.7\pm5.5$&$3.0\pm1.7$&$0.8\pm1.0$&0.84 &&SSS-$3\sigma$\\
X19& 12:39:54.1&-11:37:13.0&$4.5\pm2.3$&$4.0\pm2.0$&$2.0\pm1.4$&0.33 \\
X20& 12:39:54.6&-11:36:02.5&$5.3\pm2.4$&$1.0\pm1.0$&$2.0\pm1.5$&0.26 &&CFHT\\
X21& 12:39:54.8&-11:37:30.3&$1.5\pm1.4$&$6.1\pm2.5$&$1.0\pm1.0$&0.21 &&SSS-med\\
X22& 12:39:54.9&-11:37:08.6&$6.0\pm2.7$&$7.9\pm2.9$&$11.4\pm3.3$&0.78 \\
X23& 12:39:55.1&-11:37:37.0&$0.6\pm1.0$&$11.3\pm3.3$&$20.5\pm4.5$&1.01 \\
X24& 12:39:55.4&-11:38:48.9&$17.2\pm4.1$&$10.3\pm3.2$&$11.8\pm3.4$&1.22& &CFHT\\
X25& 12:39:55.5&-11:37:32.1&$4.7\pm2.2$&$3.9\pm2.0$&$1.9\pm1.4$&0.32 \\
X26& 12:39:55.8&-11:34:48.1&$19.4\pm4.5$&$16.3\pm4.0$&$14.9\pm3.9$&1.56 &$B=19.3$,$R=18.4$&CFHT\\
X27& 12:39:55.8&-11:37:31.4&$3.2\pm2.0$&$7.2\pm2.7$&$3.9\pm2.0$&0.42 \\
X28& 12:39:55.9&-11:37:42.1&$1.5\pm1.4$&$3.9\pm2.0$&$4.9\pm2.2$&0.32 \\
X29& 12:39:56.4&-11:31:54.2&$32.6\pm5.9$&$40.3\pm6.3$&$32.5\pm5.9$&3.07 &$B=19.7$,$R=18.3$,$K=15.6$&CFHT\\
X30& 12:39:56.4&-11:37:59.0&$17.2\pm4.1$&$10.3\pm3.2$&$6.0\pm2.4$&1.03 \\
X31& 12:39:56.4&-11:37:24.3&$13.9\pm4.1$&$11.4\pm3.4$&$6.1\pm2.4$&0.99 \\
X32& 12:39:56.5&-11:38:30.5&$7.9\pm2.8$&$5.1\pm2.2$&$10.0\pm3.1$&0.68 \\
X33& 12:39:56.9&-11:36:58.6&$3.7\pm2.0$&$3.9\pm2.0$&$8.1\pm2.8$&0.42 \\
X34& 12:39:57.2&-11:37:03.9&$5.9\pm2.5$&$1.9\pm1.4$&$2.6\pm1.7$&0.32 \\
X35& 12:39:57.2&-11:36:33.6&$4.5\pm2.2$&$3.1\pm1.7$&$0.8\pm1.0$&0.21 &&CFHT,SSS-noh\\
X36& 12:39:57.3&-11:36:29.5&$5.8\pm2.5$&$2.9\pm1.7$&$-0.2\pm0.1$&0.19 &&SSS-noh\\
X37& 12:39:57.3&-11:37:50.3&$6.1\pm2.5$&$3.1\pm1.7$&$0.0\pm0.0$&0.23 &&SSS-med\\
X38& 12:39:57.4&-11:37:19.8&$51.1\pm7.1$&$43.9\pm6.6$&$46.4\pm6.8$&4.17\\
X39& 12:39:57.4&-11:37:52.8&$14.4\pm3.8$&$12.4\pm3.5$&$7.9\pm2.8$&1.07 \\
X40& 12:39:57.4&-11:37:33.6&$5.8\pm2.5$&$3.0\pm1.7$&$5.8\pm2.5$&0.46\\
X41& 12:39:57.7&-11:38:52.9&$11.9\pm3.5$&$11.2\pm3.3$&$6.9\pm2.6$&0.93 \\
X42& 12:39:57.7&-11:37:14.6&$6.9\pm2.7$&$3.9\pm2.0$&$2.0\pm1.4$&0.40\\
X43& 12:39:57.9&-11:37:02.3&$15.9\pm4.0$&$12.2\pm3.5$&$3.0\pm1.7$&0.96 \\
X44& 12:39:57.9&-11:37:34.5&$9.0\pm3.0$&$15.4\pm3.9$&$4.8\pm2.2$&0.87 \\
X45& 12:39:58.3&-11:37:17.8&$18.3\pm4.3$&$10.3\pm3.2$&$7.0\pm2.6$&1.06 \\
X46& 12:39:58.5&-11:37:21.2&$12.1\pm3.5$&$3.9\pm2.0$&$6.1\pm2.4$&0.68 \\
X47& 12:39:58.6&-11:37:27.6&$7.0\pm2.7$&$8.4\pm2.9$&$8.0\pm2.8$&0.72 &&\hst\\
X48& 12:39:58.7&-11:38:20.6&$8.3\pm2.9$&$2.1\pm1.4$&$-0.1\pm0.0$&0.23 &&SSS-noh\\
X49& 12:39:58.7&-11:37:21.4&$8.1\pm2.8$&$7.2\pm2.7$&$3.8\pm2.0$&0.59 \\
X50& 12:39:58.8&-11:37:24.9&$27.6\pm5.3$&$13.6\pm3.6$&$11.3\pm3.3$&1.59 &&\hst\\
X51& 12:39:58.8&-11:37:00.2&$3.7\pm2.0$&$5.0\pm2.2$&$2.6\pm1.7$&0.35 \\
X52& 12:39:58.9&-11:36:54.6&$9.8\pm3.2$&$9.3\pm3.0$&$6.9\pm2.6$&0.81& &CFHT\\
X53& 12:39:58.9&-11:38:37.9&$10.8\pm3.3$&$21.4\pm4.6$&$8.0\pm2.8$&1.24& &CFHT,\hst\\
X54& 12:39:59.0&-11:35:12.0&$24.0\pm4.9$&$40.2\pm6.3$&$22.08\pm4.7$&2.62& &CFHT\\
X55& 12:39:59.1&-11:37:19.4&$61.0\pm7.8$&$28.9\pm5.4$&$10.3\pm3.2$&2.93 &&\hst\\
X56& 12:39:59.3&-11:38:28.1&$18.3\pm4.2$&$20.3\pm4.5$&$10.5\pm3.2$&1.51& &CFHT,\hst\\
X57& 12:39:59.3&-11:36:50.0&$8.7\pm3.0$&$1.8\pm1.4$&$-0.3\pm0.1$&0.25 &&SSS-med\\
X58& 12:39:59.3&-11:38:46.0&$1.8\pm1.4$&$3.0\pm1.7$&$2.9\pm1.7$&0.24& &CFHT,\hst\\
X59& 12:39:59.4&-11:37:26.8&$57.4\pm7.5$&$39.7\pm6.3$&$25.5\pm5.0$&3.78 &&\hst\\
X60&12:39:59.4&-11:37:22.7&$709.4\pm26.6$&$999.9\pm31.6$&$906.5\pm30.1$&80.32& &Nucleus\\
X61& 12:39:59.5&-11:37:01.8&$22.4\pm4.7$&$16.4\pm4.0$&$19.5\pm4.4$&1.80 \\
X62& 12:39:59.6&-11:37:29.0&$8.7\pm3.0$&$6.2\pm2.5$&$-0.3\pm0.1$&0.37 &&SSS-med\\
X63& 12:39:59.7&-11:35:24.8&$18.9\pm4.3$&$21.2\pm4.6$&$19.1\pm4.3$&1.73& &CFHT\\
X64& 12:39:59.8&-11:37:16.1&$41.3\pm6.4$&$10.1\pm3.2$&$1.8\pm1.4$&1.33 &&SSS-HR1\\
X65& 12:39:59.8&-11:34:54.4&$10.7\pm3.3$&$8.0\pm2.8$&$6.2\pm2.6$&0.77& &CFHT\\
X66& 12:39:59.8&-11:37:00.2&$7.9\pm2.8$&$9.2\pm3.0$&$4.1\pm2.0$&0.66 \\
X67& 12:39:59.9&-11:36:22.8&$36.7\pm6.1$&$37.2\pm6.1$&$22.4\pm4.7$&2.89 \\
X68& 12:39:59.9&-11:37:39.7&$2.7\pm1.7$&$2.0\pm1.4$&$3.0\pm1.7$& 0.24\\
X69& 12:40:00.0&-11:37:26.6&$11.2\pm3.3$&$11.4\pm3.3$&$7.0\pm2.6$&0.91 \\
X70& 12:40:00.0&-11:37:08.5&$28.7\pm5.3$&$35.0\pm5.9$&$19.5\pm4.4$& 2.56\\
X71& 12:40:00.1&-11:35:42.0&$13.8\pm3.7$&$-0.0\pm0.0$&$0.0\pm0.0$&0.14 &&SSS-HR\\
X72& 12:40:00.2&-11:37:22.6&$19.7\pm4.4$&$20.8\pm4.5$&$13.2\pm3.6$& 1.59\\
X73& 12:40:00.3&-11:37:23.2&$54.0\pm7.3$&$29.1\pm5.4$&$19.4\pm4.4$&3.06&&\hst \\
X74& 12:40:00.4&-11:37:17.6&$7.8\pm2.8$&$8.1\pm2.8$&$1.7\pm1.4$&0.55 \\
X75& 12:40:00.7&-11:37:12.5&$8.1\pm2.8$&$8.1\pm2.8$&$0.7\pm1.0$&0.42 &&SSS-med\\
X76& 12:40:00.7&-11:35:19.4&$31.8\pm5.6$&$27.1\pm5.2$&$25.1\pm5.0$& 2.55&&CFHT\\
X77& 12:40:00.7&-11:37:04.3&$13.2\pm3.6$&$12.3\pm3.5$&$15.4\pm3.9$&1.26 \\
X78& 12:40:00.7&-11:37:30.3&$3.5\pm2.0$&$5.2\pm2.2$&$2.9\pm1.7$&0.36 \\
X79& 12:40:00.9&-11:36:53.9&$138.3\pm11.7$&$111.7\pm10.5$&$42.2\pm6.5$& 8.87\\
X80& 12:40:01.0&-11:37:08.3&$6.5\pm2.7$&$7.1\pm2.6$&$5.1\pm2.2$&0.58& &CFHT\\
X81& 12:40:01.0&-11:37:00.9&$15.0\pm3.9$&$2.5\pm1.7$&$7.0\pm2.6$&0.69 &&\hst\\
X82& 12:40:01.1&-11:37:23.8&$391.3\pm19.7$&$81.8\pm9.0$&$5.9\pm2.4$&11.48 &&SSS-$\sigma$\\
X83& 12:40:01.3&-11:37:02.0&$3.9\pm2.0$&$6.1\pm2.4$&$2.7\pm1.7$&0.39 \\
X84& 12:40:01.3&-11:37:29.8&$9.1\pm3.0$&$10.4\pm3.2$&$16.4\pm4.0$&1.08 \\
X85& 12:40:01.4&-11:40:10.1&$2.7\pm1.7$&$3.8\pm2.0$&$1.9\pm1.4$&0.26 \\
X86& 12:40:02.0&-11:39:40.1&$71.5\pm8.4$&$0.0\pm0.0$&$-0.1\pm0.1$&1.55 &&SSS-HR\\
X87& 12:40:02.0&-11:37:07.1&$6.6\pm2.6$&$7.1\pm2.6$&$2.8\pm1.7$&0.51& &CFHT,\hst\\
X88& 12:40:02.0&-11:38:46.4&$5.1\pm2.3$&$1.9\pm1.4$&$1.8\pm1.4$&0.28 \\
X89& 12:40:02.2&-11:37:53.8&$19.6\pm4.4$&$3.9\pm2.0$&$10.2\pm3.2$&1.04 \\
X90& 12:40:02.2&-11:37:18.6&$14.9\pm3.9$&$8.9\pm3.0$&$13.1\pm3.6$& 1.08\\
X91& 12:40:02.2&-11:38:01.5&$6.7\pm2.7$&$3.7\pm2.0$&$6.1\pm2.4$&0.51& &CFHT,\hst\\
X92& 12:40:02.3&-11:37:11.7&$8.7\pm3.0$&$6.2\pm2.4$&$1.0\pm1.0$&0.40 &&SSS-med\\
X93& 12:40:02.4&-11:38:52.3&$10.0\pm3.2$&$1.9\pm1.4$&$3.9\pm2.0$&0.49 \\
X94& 12:40:02.8&-11:37:16.8&$11.9\pm3.5$&$6.1\pm2.4$&$6.0\pm2.4$&0.74 \\
X95& 12:40:03.1&-11:40:04.9&$4.8\pm2.2$&$6.1\pm2.4$&$6.9\pm2.6$&0.55 &$B=18.0$,$R=16.5$,$K=14.6$&star\\
X96& 12:40:03.6&-11:37:39.2&$3.7\pm2.0$&$7.9\pm2.9$&$11.2\pm3.3$&0.70 \\
X97& 12:40:03.7&-11:32:43.0&$13.4\pm3.9$&$11.0\pm3.3$&$13.5\pm4.0$&1.10 &$B=20.7$,$R=19.6$&star\\
X98& 12:40:03.7&-11:37:00.9&$7.4\pm2.8$&$3.0\pm1.7$&$0.8\pm1.0$& 0.28 &&SSS-med\\
X99& 12:40:04.5&-11:36:37.5&$13.9\pm3.7$&$7.4\pm2.7$&$6.0\pm2.5$&0.84 \\
X100& 12:40:04.6&-11:37:35.8&$2.5\pm1.7$&$9.1\pm3.0$&$14.3\pm3.7$&0.81 \\
X101& 12:40:04.7&-11:38:28.8&$1.2\pm1.4$&$4.0\pm2.0$&$5.1\pm2.2$&0.33 &&CFHT,\hst\\
X102& 12:40:04.8&-11:37:21.0&$15.4\pm4.0$&$2.7\pm1.7$&$1.8\pm1.4$&0.48 &&SSS-$3\sigma$\\
X103& 12:40:05.0&-11:37:32.3&$9.0\pm3.0$&$8.3\pm2.8$&$13.3\pm3.6$&0.91 \\
X104& 12:40:05.3&-11:35:00.0&$18.0\pm4.3$&$24.2\pm4.9$&$17.1\pm4.3$&1.83 &&CFHT\\
X105& 12:40:05.5&-11:39:40.5&$23.0\pm4.8$&$1.7\pm1.4$&$4.7\pm2.2$& 0.82\\
X106& 12:40:05.7&-11:37:11.2&$20.0\pm4.5$&$14.3\pm3.8$&$8.9\pm3.0$& 1.30&&\hst\\
X107& 12:40:06.9&-11:37:22.0&$4.4\pm2.2$&$3.1\pm1.7$&$1.8\pm1.4$& 0.30\\
X108& 12:40:07.0&-11:37:53.2&$20.4\pm4.5$&$18.7\pm4.3$&$21.4\pm4.6$&1.87& &CFHT,\hst\\
X109& 12:40:08.2&-11:34:46.2&$3.9\pm2.3$&$9.4\pm3.1$&$0.8\pm1.5$&0.36 &&SSS-med\\
X110& 12:40:08.4&-11:37:11.3&$10.0\pm3.2$&$3.0\pm1.7$&$1.8\pm1.4$&0.35 &&SSS-HR1\\
X111& 12:40:08.9&-11:38:17.2&$11.4\pm3.4$&$1.9\pm1.4$&$0.0\pm0.0$&0.30 &&SSS-$3\sigma$\\
X112& 12:40:09.5&-11:36:46.6&$21.5\pm4.6$&$12.3\pm3.5$&$10.1\pm3.2$&1.32 \\
X113& 12:40:09.7&-11:36:44.7&$2.9\pm1.7$&$5.1\pm2.2$&$2.9\pm1.7$&0.34& &CFHT\\
X114& 12:40:10.4&-11:36:38.7&$41.4\pm6.5$&$33.0\pm5.7$&$26.5\pm5.1$&3.01& &CFHT\\
X115& 12:40:10.8&-11:32:58.1&$22.2\pm4.9$&$18.3\pm4.3$&$23.4\pm4.9$&1.90 &$B=19.3$,$R=17.6$,$K=14.8$&CFHT\\
X116& 12:40:11.4&-11:34:42.9&$8.2\pm3.2$&$8.2\pm3.0$&$12.5\pm3.7$&0.89 \\
X117& 12:40:12.3&-11:40:05.3&$6.2\pm2.7$&$7.0\pm2.7$&$7.5\pm2.9$&0.63 \\
X118& 12:40:12.7&-11:36:30.9&$3.9\pm2.3$&$3.8\pm2.0$&$4.1\pm2.3$&0.37 \\
X119& 12:40:12.9&-11:39:03.0&$9.0\pm3.2$&$2.8\pm1.7$&$3.8\pm2.0$&0.51 \\
X120& 12:40:13.7&-11:37:52.2&$13.4\pm3.8$&$6.8\pm2.7$&$2.6\pm1.7$&0.62 \\
X121&
12:40:15.2&-11:33:07.6&$21.3\pm5.4$&$50.1\pm7.2$&$15.2\pm4.4$&2.63&
&3 USNO stars\\
X122& 12:40:15.5&-11:37:34.8&$6.1\pm2.6$&$5.5\pm2.4$&$4.8\pm2.4$&0.48 \\
\enddata

\normalsize
\tablecomments{
The
columns give the source number, the position (J2000.0), the net counts
in the three energy bands (soft: 0.1--1.1 keV; medium: 1.1--2 keV;  hard: 2--7 keV),
the 0.3--7 keV luminosity. The conversion to luminosities assumes an absorbed power-law spectrum with a photon index of
2 and $N_H = 5\times10^{20}$ cm$^{-2}$. For SSSs, we assume a
blackbody model with kT=125 eV and  $N_H = 5\times10^{20}$ cm$^{-2}$.
The $B$ and $R$ magnitudes are taken from USNO-B1.0
catalog, while the $K$ magnitude is 
from the 2MASS catalog. CFHT: Globular clusters from the ground; \hst:
Globulars clusters from \hst; SSS-HR, SSS-HR1, SSS-$3\sigma$, SSS-$3\sigma1$,
SSS-$\sigma$, SSS-med, SSS-noh: SSS category from Di\,Stefano and Kong (2003b).} 
\tablenotetext{a}{Luminosity in 0.3--7 keV (erg s$^{-1}$), assuming
$d=8.9$ Mpc, and a power-law model with $N_H=5\times10^{20}$
cm$^{-2}$, and $\alpha=2$ for all
sources except SSSs for which a blackbody model with
$N_H=5\times10^{20}$ cm$^{-2}$, and $kT=125$eV is applied.}

\end{deluxetable}

\begin{table*}
\caption{Spectral fits to the brightest X-ray sources in M 104}

\begin{tabular}{lccccc}
\hline
\hline
Object & $N_H$& $\alpha$ & kT/kT$_{RS}$ & $\chi^2_{\nu}$/dof & $L_X$$^a$\\
       &($10^{20}$cm$^{-2}$)& & (keV) & & \\
\hline
X5 &$11.1^{+10}_{-6}$ & $2.86^{+0.68}_{-0.57}$ & $0.36^{+0.05}_{-0.06}$$^b$& 1.64/25 & 26.5$^c$\\
X9 &$6.1^{+7.6}_{-6.0}$ & $2.0^{+0.45}_{-0.47}$ & & 0.8/5 & 6.34\\
X38&$2.1^{+8.4}_{-2.1}$ & $1.13^{+0.39}_{-0.32}$ & & 0.85/10 &5.89\\
X59&$5.1^{+10}_{-5.1}$ & $1.78^{+0.33}_{-0.53}$ & & 0.88/8 & 4.27\\
X79&$10.8^{+6.6}_{-6.7}$ & $2.04^{+0.43}_{-0.28}$ & & 0.83/10 & 12\\
X82&$0.9^{+4.6}_{-0.9}$ & & $0.18^{+0.02}_{-0.02}$ & 0.71/15 & 8.9\\
\hline
\end{tabular}

Note. --- All quoted uncertainties are at the 90\% confidence level.\\
$^a$ Unabsorbed luminosity ($\times10^{38}$ erg s$^{-1}$) in 0.3--7 keV,
assuming a distance of 8.9 Mpc.\\
$^b$ Raymond-Smith temperature.\\
$^c$ X5 is a foreground star. If assuming a distance of 4kpc, the
0.3--7 keV luminosity is $5.36\times10^{32}$ erg s$^{-1}$.
\end{table*}


\begin{references}

\reference{} Angelini, L., Loewenstein, M., \& Mushotzky, R. F. 2001, ApJ, 557, L35 

\reference{} Bellazzini, M., 
Pasquali, A., Federici, L., Ferraro, F.~R., \& Pecci, F.~F.\ 1995, ApJ, 
439, 687 

\reference{} Brandt, W.~N.~et al.\ 2001, \aj, 122, 2810

\reference{} 
Chiang, E.~\& 
Rappaport, S.\ 1996, \apj, 469, 255 


\reference{} Cowley, A.~P., Schmidtke, P.~C., 
Crampton, D., \& Hutchings, J.~B.\ 1990, ApJ, 350, 288

\reference{} Crawford, D. F., Jauncey, D. L., \& Murdoch, H. S. 1970, ApJ, 162, 405

\reference{} Cutri, R. M., et al. 2000, Explanatory Supplement to the 2MASS Second Incremental Data
     Release

\reference{} Davis, J.E. 2001, ApJ, 562, 575

\reference{} 
Delain, K.~M., Forman, 
W.~R., Jones, C., Murray, S.~S., \& Kraft, R.~P.\ 2001, American 
Astronomical Society Meeting, 199, 1903 


\reference{} 
Di\,Stefano, R., Paerels, F., \& Rappaport, S.\ 1995, \apj, 450, 705 



\reference{} Di\,Stefano, R., \& Rappaport, S. 1994, ApJ, 437, 733

\reference{} Di\,Stefano, R., Greiner, J., Murray, S., \& Garcia, M.\ 2001, \apjl, 551, L37 

\reference{} Di\,Stefano, R., Kong, A. K. H., Garcia, M. R., Barmby, P., Greiner, J., Murray, S. S., \& Primini, F.
     A. 2002, ApJ, 570, 618

\reference{} Di\,Stefano, R., \& Kong, A. K. H. 2003a, ApJ, in press (astro-ph/0301162)

\reference{} Di\,Stefano, R., \& Kong, A. K. H. 2003b, ApJ, in preparation

\reference{} Fabbiano, G. \& Juda, J.Z. 1997, ApJ, 476, 666

\reference{} Ford, H.C., Hui, X., Ciardullo, R., Jacoby, G.H., \&
Freeman, K.C. 1996, ApJ, 458, 455

\reference{} Forman, W.~R. et al. 2003, in preparation.  

\reference{} Friedman, R.~B., Di 
Stefano, R., Kong, A.~K.~H., Barmby, P., \& Kundu, A.\ 2002, American 
Astronomical Society Meeting, 201, 5414

\reference{} Freeman, P. E., Kashyap, V., Rosner, R., \& Lamb, D. Q. 2002, ApJS, 138, 185

\reference{} Giacconi, R.~et al.\ 2001, \apj, 551, 624

\reference{} Greiner, J., Hasinger, G., \& Kahabka, P.\ 1991, A\&A, 246, L17

\reference{}
Greiner, J., 
Tovmassian, G.~H., Di Stefano, R., Prestwich, A., Gonz{\' a}lez-Riestra, 
R., Szentasko, L., \& Chavarr{\'{\i}}a, C.\ 1999, \aap, 343, 183 

\reference{} Greiner, J.\ 2000, New Astronomy, 5, 137

\reference{}
Greiner, J.~\& 
di Stefano, R.\ 1999, Highlights in x-ray astronomy : international 
symposium in honour of Joachim Tr{\" u}mper's 65th birthday, June 17-19, 
1998, Garching, Germany : symposium proceedings / edited by Bernd 
Aschenbach {}Michael J.~Freyberg.~Garching : Max-Planck-Institut f{\" 
u}r extraterrestrische Physik, 1999.~(MPE report, No.~272, ISSN 0178-0719), 
p.66, 66 

\reference{} Harris, D. E., et al. 1993, The Einstein Observatory Catalog of IPC X-Ray
Sources. Volume 1E: Documentation (NASA TM-108401), 121

\reference{} Kong, A. K. H., Garcia, M. R., Primini, F. A., Murray,
S. S., Di Stefano, R., \& McClintock, J. 2002a, ApJ, 577, 738

\reference{} Kong, A. K. H., Garcia, M. R., Primini, F. A., \& Murray, S. S. 2002b, ApJ, 580, L125

\reference{} Kong, A.~K.~H.~\& Di\,Stefano, R.\ 2003, ApJ, 590, L13

\reference{} Kong, A.K.H., Di\,Stefano, R., Garcia, M.R., \& Greiner,
J. 2003a, ApJ, 585, 298 

\reference{} Kong, A.K.H., Sjouwerman, L.O., Williams, B.F., Garcia, M.R., \& Dickel, J.R. 2003b, ApJ, 590, L21

\reference{} Kundu, A., Maccarone, T.~J., \& Zepf, S.~E.\ 2002, ApJ, 574, L5

\reference{} Kylafis, N.~D.~\& Xilouris, E.~M.\ 1993, A\&A, 278, L43

\reference{} Larsen, S.S., Forbes, D.A., \& Brodie, J.P. 2001, MNRAS,
327, 1116

\reference{} Long, K.~S., Helfand, D.~J., \& Grabelsky, D.~A.\ 1981, ApJ, 248, 925

\reference{} Maccarone, T.~J., Kundu, A., \& Zepf, S.~E.\ 2003, \apj, 586, 814

\reference{} Monet D.G, et al. 2003, The USNO-B1.0 Catalog
 
\reference{} Pellegrini, S., Baldi, A., Fabbiano, G., \& Kim, D.-W. 2003, ApJ, submitted

\reference{} Pellegrini, S., 
Fabbiano, G., Fiore, F., Trinchieri, G., \& Antonelli, A.\ 2002, A\&A, 383, 1 
%Nuclear and global X-ray properties of LINER galaxies;
%Chandra, Beppo SAX, Sombrero and NGC 4736. 

\reference{} Pellegrini, S., 
Venturi, T., Comastri, A., Fabbiano, G., Fiore, F., Vignali, C., Morganti, 
R., \& Trinchieri, G.\ 2003, \apj, 585, 677 

\reference{} Phinney, 
E.~S.~\& Sigurdsson, S.\ 1991, \nat, 349, 220 


\reference{} 
Rappaport, S., Chiang, E., Kallman, T., \& Malina, R.\ 1994, \apj, 431, 237 

\reference{}
Remillard, R.~A., Rappaport, S., \& Macri, L.~M.\ 1995, \apj, 439, 646 



\reference{} Sarazin, C.L., Irwin, J.A., \& Bregman, J.N,  2001, ApJ, 556, 533



\reference{} Swartz, D.~A., Ghosh, 
K.~K., Suleimanov, V., Tennant, A.~F., \& Wu, K.\ 2002, \apj, 574, 382 

\reference{} Vandalfsen, M., Kavelaars, J.~J., Harris, W.~E., Hanes, D., \& Harris, G.\ 2001, Bulletin d'information du telescope Canada-France-Hawaii, 43, 10 

\reference{} van den Heuvel, E.P.J., Bhattacharya, D., Nomoto, K., \& 
 Rappaport, S.A. 1992, A\&A, 262, 97 

\reference{} Wang, Q.\ 1991, \mnras, 252, 47P 


\reference{} White, R.~E., Sarazin, C.~L., \& Kulkarni, S.~R.\ 2002, ApJ, 571, L23

\end{references}
\end{document}